# Activated Migration of Localized Ligand-Field Excitons in Atomically Thin $CrCl_3$

Hyesun Kim[1], Renlong Liu[2], Sangho Yoon[3], Hyunjong Lim[2], Takashi Taniguchi[4], Kenji Watanabe[5], Jonghwan Kim[3,6,7], Changgu Lee[2] and Sunmin Ryu[1]*

[1]Department of Chemistry, Pohang University of Science and Technology (POSTECH), Pohang, Gyeongbuk 37673, Republic of Korea

[2]School of Mechanical Engineering, Sungkyunkwan University, Suwon, Gyeonggi 16419, Republic of Korea

[3]Department of Materials Science and Engineering, Pohang University of Science and Technology (POSTECH), Pohang, Gyeongbuk 37673, Republic of Korea

[4]Research Center for Electronic and Optical Materials, National Institute for Materials Science, 1-1 Namiki, Tsukuba 305-0044, Japan

[5]Research Center for Materials Nanoarchitectonics, National Institute for Materials Science, 1-1 Namiki, Tsukuba 305-0044, Japan

[6]Department of Physics, Pohang University of Science and Technology (POSTECH), Pohang 37673, Republic of Korea

[7]Center for van der Waals Quantum Solids, Institute for Basic Science (IBS), Pohang 37673, Republic of Korea

*E-mail: sunryu@postech.ac.kr

## Abstract

Two-dimensional crystals with densely packed atoms exhibit a range of emerging properties, particularly a wide variety of excitonic behaviors. Thickness-variable layered chromium trihalides with finite surface recombination sites provide an ideal system for understanding how excitons confined in octahedral ligand fields migrate on nanometer length scales, a regime that defies conventional transport probes. In this work, we demonstrate that $Cr^{3+}$-derived photoluminescence in $CrCl_3$ is spectrally thickness-independent, but its relaxation dynamics are strongly sensitive to thickness and temperature, thereby indicating significant activated migration. A diffusion-coupled surface recombination model reveals an effective out-of-plane diffusion coefficient of $4.5 \times 10^{-6}$ $cm^2/s$ for the ligand-field excitons and a diffusion activation energy of 130 meV. A quantitative Forster analysis yields a diffusion coefficient comparable to the experimental value, supporting long-range dipolar energy transfer as a plausible mechanism for the interlayer exciton migration. The activation energy is comparable to the reorganization energy independently estimated from optical Stokes shifts, suggesting that exciton transport is coupled to local lattice relaxation. Furthermore, we show that the relaxation dynamics can be systematically tuned by either enhancing or suppressing surface recombination through controlled surface reactions or encapsulation. This work not only reveals the nanoscopic transport of localized ligand-field excitons but also establishes a spectroscopic transport probe applicable to various 2D materials.

## Introduction

Localized electronic excitations in solids, such as d-d transitions of transition metal ions confined in ligand fields (LF)[1-3] or tightly bound Frenkel states in molecular solids,[4] exhibit weak mobility unlike delocalized excitons in inorganic semiconductors. The localized excitons are pictured to undergo very limited migration through resonant transfer or activated hopping,[5-7] which stands in contrast to diffusive free-particle transport observed for the delocalized excitons.[8] Furthermore, excitations in systems with strong exciton-phonon coupling undergo vibronic reorganization due to local lattice relaxation, leading to self-trapping or polaron formation.[9] Nevertheless, even slow migration can substantially alter excited-state relaxation if excitations can be transferred to nearby emitters or quenching sites,[5-7] which are well known in triplet-energy migration in molecular solids and in concentration quenching of rare-earth phosphors.[10, 11] These observations suggest that the migration of localized excitons can affect excited-state dynamics in reduced dimensions, where boundaries often provide quenching sites. Therefore, understanding the transport behavior of localized excitons in reduced dimensions is important not only for unveiling fundamental excited-state dynamics, but also for controlling luminescence pathways, interfacial energy loss and energy-transfer efficiency in low-dimensional photonic and optoelectronic materials.

In this regard, the transport of LF excitations in atomically thin two-dimensional (2D) materials remains poorly understood, despite the high areal density of transition-metal sites that can host localized electronic excitations. Thickness-tunable 2D chromium trihalides ($CrX_3$) with dense arrays of luminescing $Cr^{3+}$ ions provide a good platform for addressing this problem because of their localized LF excitations,[1] strong exciton-phonon coupling[2, 3] and low-dimensional magnetism.[12] They exhibit $Cr^{3+}$-characteristic visible absorptions,[13] originating from LF d-d transitions found for their bulk forms.[1, 14] Near-infrared (NIR) photoluminescence (PL) with nearly thickness-independent spectral lineshapes[13, 15, 16] is consistent with localized d-d excited states rather than with delocalized excitons characteristic of transition metal dichalcogenides.[17] Resonance Raman[18] and ultrafast spectroscopy[2, 3] revealed pronounced exciton-phonon coupling and signatures of polaron formation mediated by the Jahn-Teller effect in $CrI_3$ and $CrBr_3$. Large Stokes shifts of 300–400 meV observed for both materials also indicate strong localization, which could be described within the Franck–Condon vibronic picture for molecules.[19] These characteristics suggest that localized excitations in $CrX_3$ systems may undergo thermally activated migration despite the strong localization of their emissive states.

Despite these advances, the transport and relaxation dynamics of LF excitons in 2D chromium trihalides remain to be explored. Whereas energy trapping by defects native to bulk materials affects their PL,[20] the microscopic roles of these defects in exciton transport remain unclear. The thickness-independent spectral behavior does not necessarily imply similar invariant behavior in relaxation dynamics, because the large surface-to-volume ratio of 2D crystals makes excited states highly sensitive to surfaces with interfacial coupling,[21, 22] reduced dielectric screening[23] and intrinsic[20, 24] or degradation-induced disorder.[25] Indeed, nonnegligible mobility of the LF excitations is implied by the exciton-exciton annihilation in bulk crystals[20] and defect-assisted recombination in 2D forms.[26] Exciton diffusivity in bulk $CrI_3$ was recently determined using lanthanide dopants as PL quenchers.[27] However, comprehensive understanding of exciton transport in $CrX_3$ requires addressing the structural anisotropy inherent to layered materials, which accommodate two distinct migration directions, in-plane and out-of-plane. Moreover, the task is compounded experimentally by large vibronic broadening, Stokes shifts and the consequent technical difficulty of obtaining quantitatively reliable near-infrared PL spectra, in which drastically varying detector spectral sensitivities can strongly distort spectral line shapes.[28]

In this work, we report an accelerated relaxation of LF excitons in 2D $CrCl_3$ and reveal that the thickness-dependent dynamics observed over a wide temporal range are governed by out-of-plane exciton migration to surfaces that promote nonradiative recombination. UV-ozone oxidation and encapsulation with hexagonal BN (hBN) layers further reveal that surface recombination substantially affects the relaxation dynamics. Biexponential decay of time-resolved PL (TRPL) signals is described by a two-compartment scheme, in which excitons migrate between the interior and surface regions. We describe the biexponential relaxation using a three-level kinetic model of molecular photophysics and quantify the migration using a diffusion-coupled surface recombination (DCSR) model[29] for semiconductors, which yields an interlayer diffusion coefficient of $3.0 \times 10^{-7}$ $cm^2/s$ at 77 K to $4.5 \times 10^{-6}$ $cm^2/s$ at 298 K. Activation energies from variable-temperature measurements shed light on the energy barriers for nonradiative decay and migration processes.

## Results and Discussion

***Spectrally invariant ligand-field excitons of 2D $CrCl_3$.*** $CrCl_3$ consists of honeycomb Cr layers sandwiched between two Cl layers, where each $Cr^{3+}$ ion is octahedrally coordinated by six edge-sharing Cl ligands (Fig. 1a). At room temperature, $CrCl_3$ belongs to a monoclinic C2/m ($C_{2h}^3$) space group with a van der Waals gap of 3.299 Å and unit-cell parameters a = 5.959 Å, b = 10.321 Å, c = 6.114 Å (β = 108.5°).[30] Mechanically exfoliated $CrCl_3$ flakes of various thicknesses were identified by optical contrast (Fig. S1) and atomic force microscopy (Figs. 1b and S2). The step height for monolayer (1L) was 0.64 ± 0.04 nm, which is consistent with the interlayer spacing (0.580 nm) of $CrCl_3$.[30]

The low-energy electronic transitions of bulk $CrCl_3$ are described as d-d transitions of $Cr^{3+}$ in the octahedral crystal field,[1] as shown in Fig. 1c. In Fig. 1d, the differential reflectance (DR) spectrum of 5L $CrCl_3$ shows two absorption bands at 1.68 eV (738 nm) and 2.28 eV (544 nm). Note that DR is proportional to absorptance for very thin samples supported on transparent substrates (Methods).[31, 32] The PL spectrum reveals one broad peak centered at 1.38 eV (899 nm). The two absorptions correspond to the spin-allowed $^4T_{2g} \leftarrow {}^4A_{2g}$ and $^4T_{1g} \leftarrow {}^4A_{2g}$ transitions, whereas the NIR PL arises from the $^4T_{2g} \rightarrow {}^4A_{2g}$ transition.[14] We note significant variation in the PL peak energy and lineshape in the literature.[15, 20, 26] This arises partly because the transition energy lies at the edges of the spectral range of Si CCDs or InGaAs detectors (Fig. S3a). Two apparently distinct PL spectra obtained with visible and NIR detection systems (Fig. S3b) agreed with each other (Fig. S3c) when their spectral sensitivities were corrected using a white-light source (Fig. S3d). The correction also removed lineshape distortions arising from the nonuniform spectral response of various optical components in the employed setups (Figs. S3e & S3f).

The large Stokes shift (~300 meV) between the lowest absorption and PL peaks arises from the significant displacement of the equilibrium position of $^4T_{2g}$, which was attributed to the Jahn-Teller effect for $CrI_3$.[2] The nuclear relaxation following a Franck-Condon excitation is responsible for the substantial linewidth (~250 meV) of the PL peak, which will be corroborated with temperature-dependent analysis below. As suggested by earlier reports,[15] the lineshapes of both absorption and PL showed little difference across the entire thickness range (Fig. S4). This is consistent with the picture of localized LF excitations or polaronic excitons,[2, 3] which characterize $CrX_3$ solids. Despite the invariant lineshapes, however, the steady-state PL intensities exhibited nonlinear growth unlike the absorptance with increasing thickness (Fig. 1e): the PL increase up to ~5L is 6 times slower than that for >6L. Based on the absorption and Raman signals (Fig. S4), both proportional to thickness, we tentatively attribute the reductions in PL in the first several layers to structural disorder in the surface layers. As shown in the inset of Fig. 1e, the effective PL quantum yield, defined as the PL/absorption intensity ratio, is 60% lower for the surface regions. This suggests that LF excitations undergo enhanced nonradiative relaxation, possibly at surface quenching sites, a point that will be corroborated below.

***Thickness-dependent exciton relaxation.*** Using TRPL measurements, we reveal that the relaxation of LF excitons in 2D $CrCl_3$ is strongly dependent on its thickness despite their spectral invariance. Figure 2a shows the TRPL traces generated with a 700 nm pulsed excitation beam obtained for 1–10L (see Fig. S5 for thicker samples). The PL time scale increased by one order of magnitude with increasing thickness up to 10L (Fig. 2a) and exceeded the maximum detection time window (16 μs) for thicknesses > 60L (Fig. S5). In Fig. 2b, the thickness dependence of the PL relaxation is statistically characterized by two decay time constants ($\tau_1$ and $\tau_2$) and an amplitude ratio ($A_1/A_2$), which were extracted from double-exponential fits (Fig. S6): the fast and slow time constants increased, respectively, from 90 and 600 ps for 1L to 2 and 35 ns for 18L. The tens-of-times increase in lifetime was accompanied by a rapid increase in the contribution of the slow-decay component.

To model the thickness-sensitive relaxation accompanied by thickness-invariant spectra, we devise a two-compartment scheme for 2D $CrCl_3$ slabs of thickness d, consisting of surface and interior regions (Fig. 2c). LF excitons localized at Cr sites undergo radiative or nonradiative decay, the latter being facilitated by structural defects and disorder.[26, 33, 34] Whereas the interior region has native defects, the surface has additional extrinsic defects. Additional nonradiative pathways may also arise from interface disorder or possible interfacial charge- or energy-transfer processes. These microscopic processes are not distinguished in the present work but are considered collectively as surface-related nonradiative channels. We adopt two models that allow inter-compartment migration of excitons, consequently endowing the overall relaxation with thickness dependence. In the first approach (Supplementary Note A), we employ the DCSR model, which is widely used to analyze charge-carrier lifetimes in semiconductors.[29, 35] In the current system, excitons that replace charge carriers diffuse to the surface and are quenched:[36] the transport and removal processes are characterized by the intrinsic excited-state bulk lifetime ($\tau_{bulk}$), effective diffusion coefficient ($D_{eff}$) and surface recombination velocity (S), respectively. Then, the exciton density in a thin symmetric slab (Fig. 2c) can be described with a one-dimensional diffusion-recombination equation, which generates multiple spatial modes. According to the model, the late-time TRPL signals, dominated by the fundamental spatial mode with the longest decay time, can be equated with the experimental $\tau_2$ in Fig. 2b. In addition, its decay rate ($1/\tau_2$) can be explicitly related to the thickness of the slab in low-S and high-S limiting cases: $\frac{1}{\tau_2} \approx \frac{1}{\tau_{bulk}} + \frac{2S}{d}$ and $\frac{1}{\tau_2} \approx \frac{1}{\tau_{bulk}} + \frac{\pi^2 D_{eff}}{d^2}$, respectively (Supplementary Note A). As shown in Fig. 2d, the PL decay rate scales linearly with $1/d^2$, rather than $1/d$ (Fig. S7). This indicates that the diffusion-recombination process operates in the high-S limit, where surface recombination is highly efficient and diffusion to the surface limits the overall relaxation dynamics. $D_{eff}$ obtained from the slope is $4.5 \times 10^{-6}$ $cm^2/s$, several orders of magnitude smaller than that of delocalized excitons.[8] This comparison verifies the strongly localized nature of the LF excitons, which will be further characterized by their temperature dependence below.

The second model, adapted from a three-level system commonly used in molecular photophysics[25, 37], assumes that the $^4T_{2g}$ states of both regions are degenerate and exchange populations via generalized diffusion at a rate constant $k_{ex}$ (Supplementary Note B). Defining radiative and nonradiative decay channels for each region, the TRPL trace of the slab is given as the sum of two exponential decays, which essentially correspond to $\tau_1$ and $\tau_2$ in Fig. 2b. This minimal model illustrates how population exchange between the surface and interior regions can give rise to the experimentally observed biexponential PL decay, although it does not quantitatively reproduce the distinct thickness dependences of the two decay components without additional assumptions on the rate constants. Thus, the three-level model provides a kinetic interpretation of the biexponential relaxation, whereas the DCSR model quantitatively describes its thickness dependence and yields the effective diffusion coefficient.

***Modulation of surface recombination.*** The finding that exfoliated $CrCl_3$ samples lie in the high-S limit (Fig. 2d) suggests that their surfaces contain a substantial number of extrinsic defects. They may be generated by mechanical exfoliation, exposure to air, or laser light, which are required for PL measurements. As shown for $CrI_3$,[38, 39] native surface defects in $CrCl_3$[40] are likely generated by a photocatalytic aquation or oxidation reaction, which substitutes halide ligands of photoexcited $CrX_3$ with $H_2O$ or oxygen. Using a time-lapse control experiment, however, we verified that the initial 8 h of exposure to air, or repeated PL measurements under ambient conditions, did not alter PL intensity or decay kinetics (Fig. S8). This suggests that the majority of extrinsic surface defects are generated in the very early stage of exposure to air and light. To provide a chemical basis for the conclusions from the diffusion-recombination analysis, we investigated how surface modification affects exciton relaxation.

First, oxidation by UV-generated ozone (UVO) was used to induce additional surface defects.[25] With an increase in UVO exposure to 5 s, the PL intensity of 6L decreased to one-third, whereas the lineshape was maintained (Fig. 3a). Both PL decay times were shortened by half for the same UVO treatment, which also increased the contribution of the fast component significantly (Fig. 3b). These observations indicate that the surface oxidation primarily enhances nonradiative decay near the surface, equivalently increasing S, but not creating distinct PL emitters. Whereas a similar observation was made for a wide range of thicknesses (Fig. 3c), the effects of the surface reactions decreased with increasing thickness. This agrees with the $1/d^2$ scaling in the high-S limit in that the contribution of the diffusion-recombination channel is smaller for thicker samples.

Next, we encapsulated exfoliated $CrCl_3$ crystals with thin hBN flakes to reduce ambient or photoinduced degradation.[41] In one approach, $CrCl_3$ was partially encapsulated by dry-transferring it onto hBN/quartz to avoid contact with interfacial water and air that linger on hydrophilic quartz substrates.[22, 42] In another approach, $CrCl_3$ was exfoliated and sandwiched between two hBN flakes in an $N_2$-filled glove box to minimize ambient degradation.[38] As shown in Fig. 3c, $CrCl_3$/hBN/quartz exhibited a 4-fold increase in $\tau_2$, indicating that the S value is significantly lower than that of pristine samples. A DCSR analysis of the on-hBN data using the asymmetric boundary condition (passivated bottom, exposed top; Equation A9 in Supplementary Note A) yields $D_{eff} \approx 4.5 \times 10^{-6}$ cm$^2$/s, in agreement with the on-quartz value. This result indicates that the lifetime enhancement arises primarily from the modified boundary condition rather than from a change in the intrinsic out-of-plane diffusion coefficient. Moreover, the hBN-sandwiched samples showed even longer lifetimes. The decrease in S in both cases indicates that the hBN layers protect $CrCl_3$ from photoaquation or photooxidation by blocking $H_2O$ or $O_2$.[38, 39] Notably, exciton relaxation can be slowed by an order of magnitude when both surfaces are encapsulated in an inert gas atmosphere, consistent with the diffusion-recombination model.

***Thermally activated relaxation and diffusion.*** To shed more light on the nature of the localized excitons, we investigated their temperature dependence. Figure 4a shows that the localized $Cr^{3+}$ emission is 5 times decreased, accompanied by a substantial redshift and broadening upon heating from 4 to 290 K. The PL attenuation at higher temperature can be attributed to activated nonradiative decay. This may arise from enhanced vibronic coupling between LF-excited states and lattice vibrations, as described in classical vibronic relaxation theories of $Cr^{3+}$ systems.[33, 34] In Fig. 4b, the temperature dependence of the PL intensity, fitted with a standard Arrhenius plot, suggests the presence of two activation barriers (20 and 4 meV). In an LF picture, the higher-energy barrier can be attributed to vibrations that promote nonradiative vibronic relaxation of the LF excitons. Stretching and bending modes of X-Cr-X, strongly coupled to the polaron formation in $CrI_3$,[2] may be responsible for the nonradiative vibronic coupling. Whereas the lower-energy channel may be associated with nonradiative vibronic relaxation facilitated at disordered LF sites on surfaces,[33] a clear identification requires further investigation.

Thermal energy also affects the spectral lineshape through thermally populated vibrational states of the emitting $^4T_{2g}$ level. The resulting increase in lattice fluctuations leads to broadening of the emission band, which can be described by the following model function for the temperature-dependent FWHM:[27] $\mathrm{FWHM}(T) = \mathrm{FWHM}(0)\sqrt{\coth(\frac{h\nu_{eff}}{2k_B T})}$, where $\nu_{eff}$ corresponds to the effective normal mode responsible for the fluctuation. As shown in Fig. 4c, the fit reveals $\nu_{eff} = 203 \pm 10$ $cm^{-1}$, which can be associated with $A_g$ or $B_g$ Raman modes (Fig. S4c).[43] Whereas the peak shift, phenomenologically fitted with the Varshni equation,[44] may also be affected by electron-phonon coupling, quantitative analysis is not attempted due to possible contribution of substrate-enforced thermal expansion, which is observed for 2D materials.[45, 46] In this regard, we note that bulk behavior of $CrI_3$[47] and $CrBr_3$[48] are consistent with that of 2D $CrCl_3$ shown in the current work.

Exciton diffusion to surface quenching sites provides an additional origin for the temperature dependence in PL. As shown in Fig. 4d, the TRPL decay rates accelerate as the temperature increases from 77 to 298 K. Quantitatively, the slow decay time constant decreases by a factor of 12–15 across a wide range of thicknesses (Fig. 4e). We note that a description with more than three exponential functions is required for thickness ≥ 10L below 150 K. Within the DCSR model (Supplementary Note A), the increased number of observed decay components can be attributed to temporal separation among low-order spatial modes, although a trap-related contribution cannot be excluded. With increasing temperature, $D_{eff}$ estimated from the DCSR analysis (Fig. 4e) shows a 15-fold increase ($3.0 \times 10^{-7}$ $cm^2/s$ at 77 K to $4.5 \times 10^{-6}$ $cm^2/s$ at 298 K). An Arrhenius analysis in Fig. 4f reveals two apparent activation energies, 130 and 10 meV. The larger energy barrier is comparable in magnitude to the lattice reorganization energy ($\lambda \approx 150$ meV), which corresponds to half of the Stokes shift (~300 meV) under the harmonic configurational coordinate approximation.[20] Huang-Rhys factor[49] of ~6, estimated by assuming that $\nu_{eff}$ is responsible for the reorganization, also indicates a strong lattice relaxation leading to localization.[13] This energetic correspondence and a large Huang-Rhys factor suggest that the interlayer diffusion of LF excitons is strongly coupled to local lattice dynamics, the mechanism of which will be discussed below. We note that the smaller energy barrier likely reflects low-energy phonon-mediated or shallow-trap-mediated migration. Although these activation energies are phenomenologically derived, they capture the activated diffusion of strongly localized LF excitons depicted in Fig. 2.

Lastly, our data reveal essential details of how the excitons migrate in $CrCl_3$. The exciton diffusion described in the current work is distinct from the Wannier-type in conventional semiconductors. The optical data instead indicate a strongly localized excited state, as evidenced by the large Stokes shift and reorganization energy. Notably, the larger activation energy for diffusion is comparable to the estimated reorganization energy, suggesting exciton migration is coupled to lattice relaxation. Exciton transport in $CrI_3$ has recently been described in terms of Dexter-type hopping, where diffusion coefficients on the order of $10^{-7}$ $cm^2/s$ were reported.[27] Whereas the localized character of $CrCl_3$ excitons is apparently consistent with such a picture, there are a few notable differences. Most of all, the transport in the current observation occurs across layers and thus is not mediated by mechanisms that require wavefunction overlap, including Dexter transfer.[6] Whereas Cr ions in neighboring layers could be coupled via two Cl atoms that are separated by a van der Waals gap like a super-superexchange interaction,[50] the coupling would be far weaker than the superexchange-type coupling that is responsible for the observed in-plane diffusion in $CrI_3$.[27] These considerations suggest that a purely Dexter-type mechanism is unlikely to account for the observed transport.

On the other hand, the observed transport can be reasonably explained by a conventional relaxed-state point-dipole Forster picture. Our quantitative analysis (Supplementary Note C), using the experimentally determined donor's PL quantum yield (Supplementary Note D), acceptor absorptivity, and donor's PL and acceptor's absorption spectra, gives a Forster radius of $R_0 \approx 7.0$ Å, which is comparable to the interlayer nearest-neighbor Cr–Cr distance ($r_0 = 6.12$ Å). Whereas nearest-neighbor transfer alone gives an out-of-plane diffusion coefficient of $7.0 \times 10^{-7}$ $cm^2/s$, including transfer to Cr sites in more distant neighboring layers increases the calculated value to $5.4 \times 10^{-6}$ $cm^2/s$, comparable to the experimentally observed value of $4.5 \times 10^{-6}$ $cm^2/s$. Thus, point-dipole FRET between relaxed LF excitons is quantitatively plausible for the observed interlayer transport. Overall, the pronounced temperature dependence suggests that the transfer is strongly coupled to lattice dynamics, for which exciton-phonon coupling may play an essential role. Thermally activated lattice vibrations can enhance the FRET rate from vibronically broadened excited states by increasing the effective spectral overlap.[5] The spectral overlap required for the Forster mechanism also appears in Dexter's transfer rate. In this regard, we note that $CrI_3$ gains nonnegligible spectral overlap above 100 K.[27] Consequently, the long-range mechanism is more plausible for $CrCl_3$ than for $CrI_3$ because the smaller Stokes shift and larger PL linewidth of $CrCl_3$ are expected to increase spectral overlap and thus favor energy transfer. Furthermore, the experimentally observed activation energy (~130 meV) is comparable to the independently estimated reorganization energy (~150 meV), suggesting an important role of lattice reorganization in modulating the interlayer transfer. Whereas these observations do not require thermal detrapping from self-trapped excitons or transfer through a less-relaxed state, they are consistent with long-range energy transfer modulated by exciton-phonon coupling.

## Conclusion

We investigated the energetics, relaxation and nanometer-scale transport of excitons localized at $Cr^{3+}$ ions under an octahedral ligand field using time-resolved PL spectroscopy. By employing two types of array detectors, optimized for the visible and NIR ranges, and verifying their agreement after spectral correction with a white-light source, we recovered an undistorted PL lineshape that is significantly different from that previously reported. Whereas the emission is spectrally thickness-independent, its intensity and relaxation were highly sensitive to thickness and temperature. From the temperature dependence of PL intensity, we identified an energy barrier for nonradiative relaxation, which can be associated with lattice vibrational modes. By analyzing the thickness- and temperature-dependent dynamics with the DCSR model, we determined an effective exciton diffusion coefficient of $4.5 \times 10^{-6}$ $cm^2/s$ along the out-of-plane direction and a diffusion activation energy of 130 meV. A quantitative point-dipole Forster analysis yields an interlayer diffusion coefficient of 5.4 × $10^{-6}$ $cm^2/s$, comparable to the experimental value supporting long-range dipolar energy transfer as a plausible mechanism for the interlayer migration of LF excitons. The diffusion activation energy is comparable to the reorganization energy estimated from optical Stokes shifts, which connects exciton transport to local lattice relaxation and suggests that the efficiency of the energy transfer is modulated by exciton-phonon coupling. We also demonstrated that exciton relaxation dynamics can be modulated by either enhancing or suppressing surface recombination through controlled oxidation with UV-ozone or encapsulation with hexagonal BN. These results provide direct insight into the transport of localized ligand-field excitons in a layered magnetic crystal and establish PL dynamics as a nanoscopic transport probe applicable to various 2D materials.

## Methods

***Synthesis of bulk crystals.*** $CrCl_3$ crystals were synthesized via chemical vapor transport.[51] High-purity anhydrous $CrCl_3$ powder was loaded into a cleaned, pre-baked quartz ampoule, together with a small amount of iodine ($I_2$) as the transport agent in an inert atmosphere. The ampoule was evacuated to high vacuum and sealed. The sealed ampoule was placed in a two-zone furnace with a temperature gradient, where the source zone was maintained at 650–700 °C and the growth zone at 550–600 °C. Under these conditions, volatile chromium chloride species were transported from the hot zone to the cooler zone, where crystallization occurred over approximately 7 days. After growth, the furnace was slowly cooled to room temperature. The ampoule was then opened in an inert atmosphere, and the resulting $CrCl_3$ crystals were collected from the growth zone for further characterization.

***Preparation and characterization of 2D samples.*** 2D $CrCl_3$ samples of various thicknesses were prepared on amorphous quartz substrates or Si wafers with 85-nm $SiO_2$ using the mechanical exfoliation method.[22, 42, 52] Samples for optical measurements were prepared on quartz substrates to minimize optical interference.[53] To minimize degradation occurring in ambient conditions, prepared samples were stored in a dark vacuum desiccator maintained below 25 Torr. All optical measurements were completed within 8 h of exfoliation, during which little change was observed in PL spectra and transients (Fig. S8). The thickness of the samples was determined using optical contrast and atomic force microscopy (AFM). The optical contrast, defined as the fractional change in blue- and red-channel optical micrographs, yielded the best sensitivity and reliability for resolving thickness in samples supported on quartz and $SiO_2$/Si substrates, respectively. Height and phase AFM images were obtained with a commercial unit (Park Systems, XE-70) in the non-contact mode with silicon tips having a tip radius of 8 nm (MicroMasch, NSC-15). For single-sided encapsulation with hBN, 2D $CrCl_3$ was transferred onto hBN flakes supported on quartz substrates using a standard dry-transfer method.[42] For the most comprehensive encapsulation, some samples were exfoliated and sandwiched with two hBN flakes in a glove box using a standard pickup transfer.[54] The thickness of hBN flakes was in the range of 19–25 nm for the double-sided encapsulation and beyond 25 nm for the single-sided encapsulation.

***Steady-state PL measurements.*** Steady-state PL measurements were performed using three micro-PL setups.[22, 25, 28] Excitation beams of 632.8 nm HeNe lasers were focused onto a spot (~1 μm) of samples using an objective lens (40×, numerical aperture = 0.60). Back-scattered PL signals were collected with the same objective and directed to Czerny-Turner spectrographs (focal length of 30 cm) coupled to photodetectors: Si CCDs (Andor, Newton EM; Princeton Instruments, PyLoN) and InGaAs photodiode arrays (Andor, iDus InGaAs; Princeton Instruments, PyLoN-IR). Whereas the PL peak of $CrCl_3$ is located at the spectral edges of both detector types, the InGaAs detectors exhibit a flatter spectral response over the range of interest (Fig. S3). Thus, PL spectra recorded with the InGaAs detectors were shown unless noted otherwise. For variable-temperature measurements, samples were placed inside cryostats (Oxford Instruments, MicrostatHires; Montana Instruments, Cryostation s50) operated with liquid nitrogen ($LN_2$, 77–300 K) or liquid helium (LHe, 4–300 K). Uneven spectral responses of the detectors and other optical components in the setups, generating apparent peak shifts and shoulder-like features, were corrected using a white-light source (tungsten lamp), as described in Fig. S3. PL spectra were not rescaled with a Jacobian transformation, which resulted in only minor changes (Fig. S9). The average excitation power was maintained below 2 μW to avoid unwanted photoinduced degradation. All measurements were performed under ambient conditions except for the variable-temperature measurements.

***Time-resolved PL measurements.*** Temporal information of the PL signals was obtained with a time-correlated single photon counting (TCSPC) device (PicoQuant, PicoHarp 300).[25] As an excitation source, a Ti:sapphire laser (Coherent, Chameleon Ultra II) with a 140 fs pulse width was tuned to 1.77 eV (700 nm) and focused to a 1.6 μm spot (FWHM) using a 40× objective lens (numerical aperture = 0.60). Back-scattered PL signals collected by the same objective lens were filtered with a long-pass filter (cutoff at 750 nm). Due to the wide variation in PL signal time spans, the repetition rate was reduced from 80 MHz to a range of 4 MHz to 62.5 kHz using a pulse selector (APE, HP-Ti:Sa). The start and stop signals for the TCSPC device were obtained from a PIN photodiode (PicoQuant, TDA200) and a single-photon avalanche diode (SPAD; Micro Photon Devices), respectively. The temporal resolution defined as the FWHM of the instrumental response function (IRF) was 50 ps at 1.77 eV. The average power was maintained as low as possible (e.g., 2 μW at 4 MHz) to avoid unwanted photoinduced degradation.

***Differential reflectance measurements.*** Differential reflectance (DR) spectra were obtained with one of the micro-PL setups using a CCD and a broadband light beam from a tungsten-halogen lamp.[55] Under a thin-film approximation,[31] DR is proportional to absorptance: DR is defined as $(R_S - R_0)/R_0$, where $R_S$ and $R_0$ are reflectance of the sample and bare transparent substrate, respectively.

ASSOCIATED CONTENT

**Supporting Information.**

Thickness dependence of optical contrast of 2D $CrCl_3$, topographic images and height profiles of atomically thin $CrCl_3$ flakes, spectral-response correction of PL spectra, thickness dependence of PL, DR and Raman spectra of 2D $CrCl_3$, TRPL data across a broad thickness range, residual analysis of TRPL decay fits, DCSR evaluation of thickness scaling of slow decay components, photostability of 2D $CrCl_3$ in ambient air, influence of Jacobian factor, three-level kinetic model for biexponential PL decay, optical spectra used for the relative PLQY estimation; Experimental quantities used for the relative PLQY estimation; Diffusion-coupled surface recombination (DCSR) model for TRPL analysis, three-level kinetic model for biexponential PL decay, point-dipole Forster rate for interlayer exciton transfer, room-temperature PL quantum yield of few-layer $CrCl_3$.

AUTHOR INFORMATION

**Corresponding Author**

*E-mail: sunryu@postech.ac.kr

NOTES

**Conflicts of Interest**

The authors declare no conflict of interest.

**Data Availability Statement**

The data that support the findings of this study are available from the corresponding author upon reasonable request.

ACKNOWLEDGMENTS

This work was supported by the National Research Foundation of Korea (NRF-RS-2024-00336324, NRF-RS-2024-00411134, 2021R1A6A1A10042944) and the Glocal University 30 project.

## FIGURE CAPTIONS

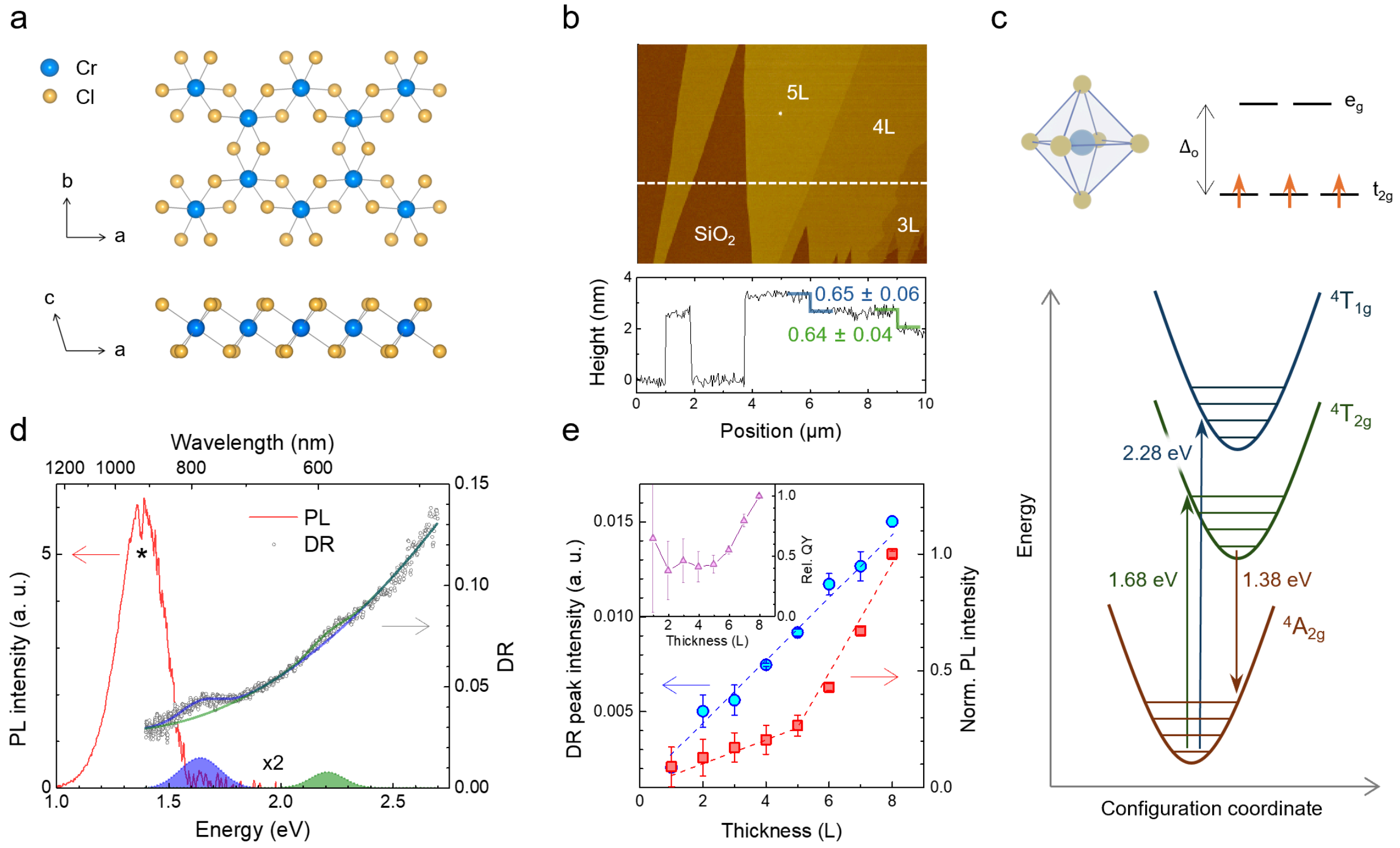

**Figure 1. Spectrally invariant LF excitons of 2D $CrCl_3$.** (a) Crystal structure of $CrCl_3$ in a-b plane (top) and a-c plane (bottom). (b) AFM height image (top) and profile (bottom) of mechanically exfoliated several-layer $CrCl_3$ on $SiO_2$/Si substrate. The height profile was obtained along the dashed line in the image. (c) Ground-state electron configuration of $Cr^{3+}$ under an octahedral ligand field generating splitting energy ($\Delta_o$) (top) and energy level diagram showing the $^4A_{2g}$ ground state, $^4T_{2g}$ and $^4T_{1g}$ excited states (bottom). The transitions observed in the absorption and PL spectra are marked with arrows. (d) PL and differential reflectance (DR) spectra of 5L $CrCl_3$. The PL spectrum, corrected for the spectral sensitivities of the detection system (Methods and Fig. S3), is dominated by a single peak at 1.38 eV. The asterisk denotes an artifact arising from incomplete spectral-response correction. The DR spectrum shows two absorption peaks at 1.68 and 2.28 eV, both well fitted with Gaussian functions. The Gaussian fits are also shown separately at 2× magnification. The broad background contribution to the DR spectrum was approximated by a quadratic function. (e) Intensity of the DR peak at 1.68 eV and PL intensity as a function of thickness. The inset shows the relative quantum yield, defined as the PL-to-DR

intensity ratio. The PL intensity and relative quantum yield are normalized to the 8L values.

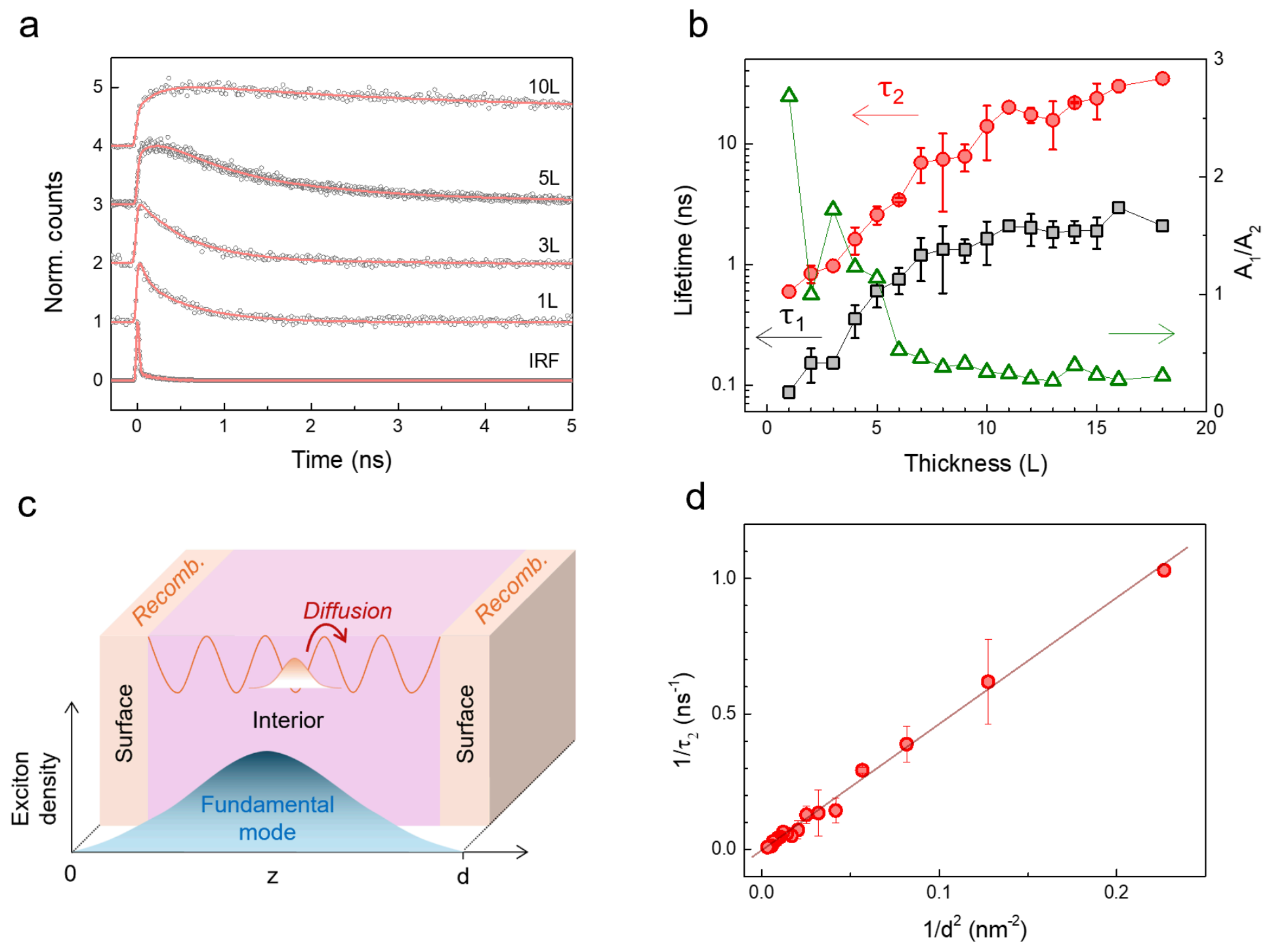

**Figure 2. Thickness-dependent PL decay originating from diffusion and surface recombination.** (a) TRPL signals of representative samples (1–10L). Data (circles) are fitted with a biexponential function (red line). The FWHM of IRF (bottom) is 50 ps. (b) Lifetimes ($\tau_1$ and $\tau_2$) and amplitude ratio ($A_1/A_2$) obtained from the biexponential fitting in (a). (c) Two-compartment scheme representing diffusion along the z-axis in the interior and nonradiative recombination (denoted as Recomb.) at the surface with surface recombination velocity (S) in a $CrCl_3$ slab of thickness d. According to the diffusion-coupled surface recombination (DCSR) model (Supplementary Note A), the spatial profile of the exciton density is reduced to the fundamental mode (solid curve) in the late time limit. (d) $1/\tau_2$ plotted against $1/d^2$ for 3–18L samples. From the slope corresponding to $\pi^2 D_{eff}$ in the high-S limit of the DCSR model, $D_{eff}$ was obtained as $4.5 \times 10^{-6}$ cm$^2$/s.

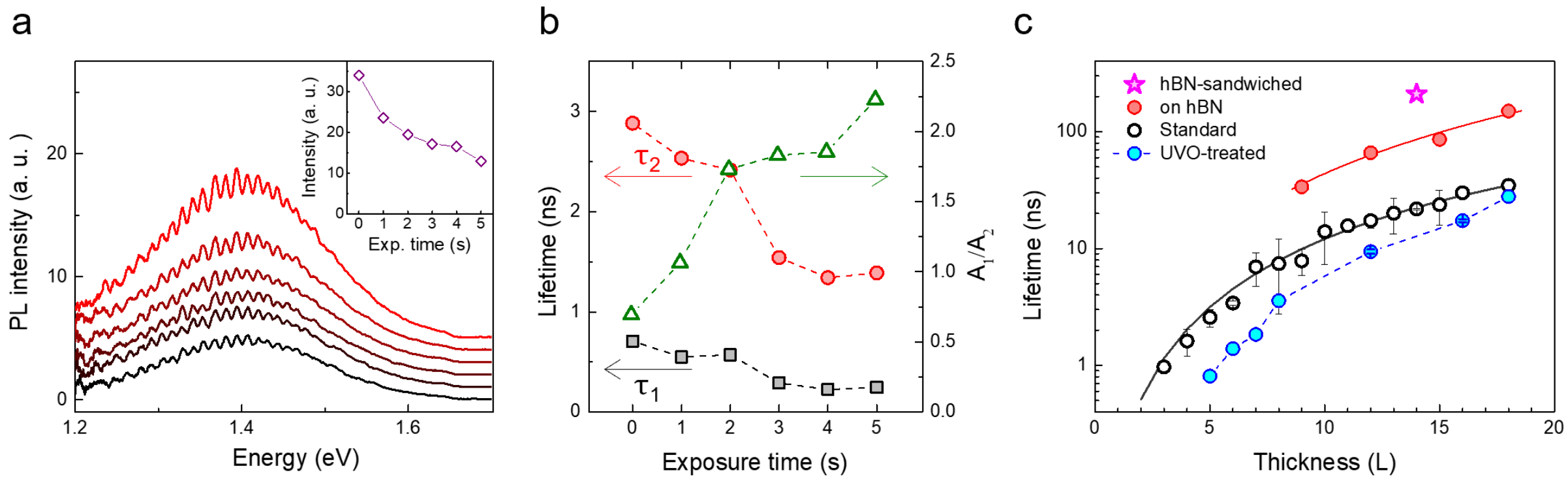

**Figure 3. Modulation of surface recombination via reaction and encapsulation.** (a) PL spectra (top to bottom) of a 6L sample obtained with increasing UVO exposure time. The inset shows the PL intensity as a function of the exposure time. (b) PL lifetimes ($\tau_1$ and $\tau_2$) and amplitude ratio ($A_1/A_2$) as a function of the exposure time. (c) Slowest lifetime ($\tau_2$ or $\tau_3$) of $CrCl_3$/quartz (Standard), $CrCl_3$/quartz UVO-treated for 5 s (UVO-treated), $CrCl_3$/hBN/quartz (on hBN) and hBN/$CrCl_3$/hBN/quartz (hBN-sandwiched). The solid lines are DCSR fits to the data; the dotted line is a guide to the eye.

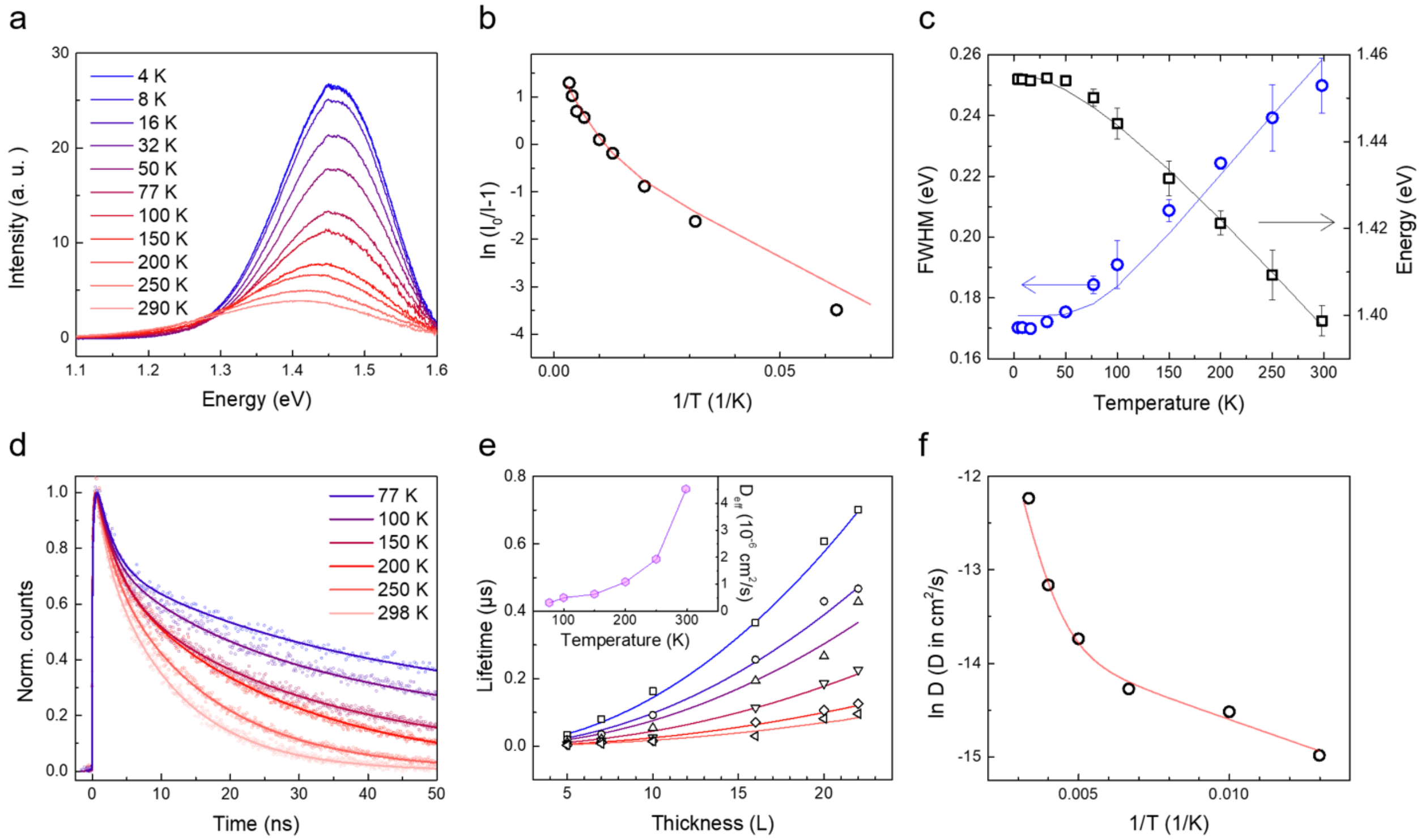

**Figure 4. Thermally activated relaxation and diffusion of LF excitons.** (a) PL spectra of 10L $CrCl_3$ obtained at 4–290 K. (b) Arrhenius analysis of PL intensity. Bilinear fit yields effective activation energies of 20 and 4 meV. (c) FWHM (left ordinate) and peak energy (right ordinate) obtained from (a). Both data (symbols) were fitted with phenomenological models (solid lines), respectively. (d) TRPL signals of 8L $CrCl_3$ at selected temperatures (77–298 K). Solid lines are multi-exponential fits to the data. (e) Slowest lifetime ($\tau_2$–$\tau_4$) obtained from TRPL measurements over a range of thickness and temperature. The inset shows $D_{eff}$ obtained from DCSR fits (solid line) in (e). (f) Arrhenius analysis of $D_{eff}$ extracted from the DCSR fits in (e). The bilinear fit yields activation energies of 130 and 10 meV.

**TOC Figure**

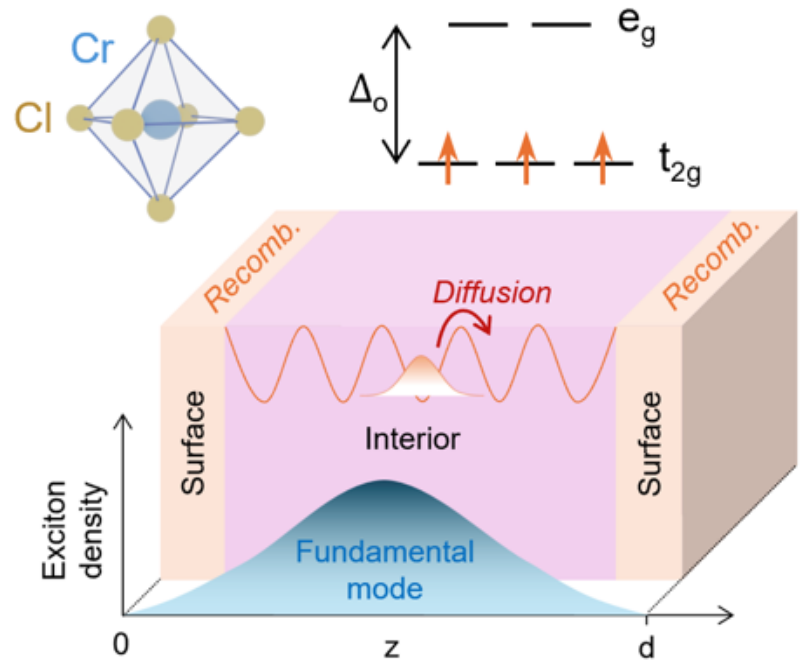

Cr
Cl
$\Delta_o$
$e_g$
$t_{2g}$
Recomb.
Recomb.
Diffusion
Surface
Surface
Interior
Exciton density
Fundamental mode
0
z
d